# Dislocation-ledge coupling governs semicoherent precipitate growth

Jin-Yu Zhang[1,2,3*], Juan Du[1,4], Lin Yang[5], Frédéric Mompiou[4], Shigenobu Ogata[2*], Wen-Zheng Zhang[1*]

[1] Key Laboratory of Advanced Materials (MOE), School of Materials Science and Engineering, Tsinghua University, Beijing 100084, China.

[2] Department of Mechanical Science and Bioengineering, The University of Osaka, Osaka 560-8531, Japan.

[3] Department of Physics, McGill University, Montréal, Québec H3A 2T8, Canada.

[4] Université de Toulouse, CNRS, CEMES, Toulouse 31055, France.

[5] School of Materials Science and Engineering, Beijing Institute of Technology, Beijing 100081, China.

* Corresponding authors:
Jin-Yu Zhang, Email address: jinyu.zhang@tsme.me.es.osaka-u.ac.jp
Shigenobu Ogata, Email address: ogata@me.es.osaka-u.ac.jp
Wen-Zheng Zhang, Email address: zhangwz@tsinghua.edu.cn

## Abstract

Many crystalline materials acquire their properties as one crystal phase grows through another, but a long-standing defect-kinetic puzzle remains: how dense interfacial dislocation networks between two crystals advance, reorganize, and accommodate strain without relying on ordinary glide. Here, using in situ transmission electron microscopy, O-lattice analysis, and three-dimensional phase-field crystal simulations, we identify how semicoherent interfaces overcome this constraint during lath precipitate growth. In situ observations of austenite precipitates in duplex stainless steel reveal nanometer-high growth ledges propagating laterally along migrating habit planes. O-lattice analysis shows how lattice misfit prescribes a closed network across habit planes, side facets, and end faces. Simulations resolve the hidden three-dimensional dynamics: the network undergoes diffusion-assisted, non-conservative motion, producing steady end-face advance and ledge-mediated broad-facet migration while accommodating transformation strain. These results reveal a general defect-kinetic route linking point-defect transport, dislocation-network motion, interface migration, and morphology evolution.

## Introduction

Many crystalline materials evolve by the advance of boundaries between lattice-mismatched crystals, from solid-state transformations in structural alloys to solid–gas transformation fronts and lattice-mismatched films (*1-6*). Such boundaries often become semicoherent: they preserve a crystallographic relationship, but carry dense dislocation networks that must move while accommodating transformation strain. The long-standing puzzle is how these networks can make a three-dimensional (3D) interface move when their required trajectories are generally not ordinary slip planes. This missing defect-kinetic link limits our ability to predict microstructure evolution and morphology selection in crystalline materials.

This problem is particularly acute for lath-, plate-, and needle-shaped precipitates in structural alloys. Such precipitates are central to precipitation strengthening, where precipitate size, morphology, volume fraction, and spatial distribution control mechanical response and microstructural stability during aging (*7, 8*). In technologically important systems, including steels, Ti alloys, and Zr alloys, precipitate–matrix interfaces are often semicoherent and contain dense interfacial dislocation arrays or networks (*9-15*). These precipitates grow highly anisotropically, with rapid lengthening but much slower thickening (*16*), producing Widmanstätten-type morphologies (*17*). Although this anisotropy has long been associated with crystallography and interface defects (*4-6*), a predictive kinetic picture linking interfacial dislocation motion, interface migration modes, and 3D precipitate growth remains missing.

Experiments have established that semicoherent interfaces bounding lath precipitates are structurally heterogeneous (*4-6*), with distinct dislocation structures on different interface regions. Broad habit planes and side facets commonly contain dislocation arrays aligned with the precipitate long axis (*10-14, 18-20*), whereas end faces, approximately normal to the long axis, often exhibit curved geometries and more complex dislocation networks (*9, 21*). Nanometer-high growth ledges (*22*) have been reported on semicoherent facets in steels (*10*), Ni/Cu-based alloys (*18, 23-26*), and Ti/Zr-based alloys (*13-15, 27*), and are widely invoked to explain stepwise facet advance. Transmission electron microscopy (TEM) (*14, 25, 27*), high-resolution TEM (*10, 13*), and crystallographic analyses (*28, 29*) have resolved ledge character and associated dislocation content in several systems. In the duplex stainless steel system examined here, representative γ-austenite precipitates in an α-ferrite matrix exhibit these characteristic interface regions and interfacial line contrasts (Fig. 1A). However, most of these observations are static or local. Much of the kinetic evidence is indirect, such as ledge-position changes after short aging treatments (*18*), and only a few in situ studies have reported lateral ledge motion on facets (*28*). Several studies have further suggested that ledge-mediated migration may involve non-conservative defect processes (*13, 24, 28*), but how these processes operate at the level of interfacial dislocations and accommodate transformation strain remains poorly understood. More broadly, it remains unclear how habit planes, side facets, and end faces operate as one coupled, evolving defect system, rather than migrating as independent interface regions, during 3D precipitate growth.

Directly resolving these non-conservative defect kinetics across semicoherent precipitate interfaces remains experimentally challenging. Recent in situ atomic-resolution studies of moving crystalline interfaces, such as grain boundaries (*1, 30-33*), segregation-induced heterointerfaces (*34*), and oxide/metal transformation fronts (*2*), have shown that interface migration can involve point-defect-assisted climb, disconnection motion, and other non-conservative defect processes. Analogous non-conservative defect kinetics may operate in semicoherent interface migrations during precipitate growth, but they are difficult to observe directly. Atomic-scale imaging must be maintained at elevated temperature, while the 3D precipitate geometry and crystallographically complex interfaces hinder alignment of interfacial defects along suitable zone axes. Dense dislocation arrays with nanometer-scale spacing can produce weak or overlapping contrast during migration, and thin-foil free surfaces may bias defect behavior relative to bulk conditions (*35*). These limitations motivate complementary approaches that can connect existing experimental observations to the hidden 3D defect dynamics of the moving interfaces.

Atomistic simulations provide alternative access to such defect dynamics, but most methods cannot simultaneously capture diffusive time scales, non-conservative dislocation motion, and sustained 3D growth of semicoherent precipitates. Molecular dynamics can resolve conservative interface migration coupled to dislocation glide (*36-38*), but its time scale limits direct treatment of diffusion-controlled processes (*39*). Monte Carlo-based approaches (*36, 40*), including grand-canonical variants (*30*), can incorporate diffusion or open-system exchange, but collective reorganization of dense interfacial dislocation networks during 3D growth remains computationally challenging. Phase-field crystal (PFC) models (*41*) and related amplitude-expansion formulations (*42*) evolve on diffusive time scales while retaining atomic-scale crystalline order and defect structures (*32, 43*). Combined with O-lattice analysis (*44*), which links lattice misfit to the geometry of interfacial dislocations, PFC simulations provide an attractive route to resolve the defect kinetics of semicoherent interface migration and 3D precipitate growth.

Here, we combine in situ TEM, 3D PFC simulations, and O-lattice analysis to identify how semicoherent crystalline interfaces advance through non-conservative dislocation-network motion. Fig. 1 summarizes representative experimental observations and the mechanism resolved in this work. In situ TEM of austenite precipitates in duplex stainless steel captures nanometer-high growth ledges propagating laterally along migrating habit planes (Fig. 1B). PFC simulations of a model body-centered-cubic (BCC) to face-centered-cubic (FCC) transformation then reveal the hidden 3D dynamics: interfacial dislocations on habit planes, side facets, and end faces remain connected as closed loops, whose diffusion-assisted, non-conservative motion produces ledge-mediated migration of broad facets and steady end-face advance along the precipitate long axis (Fig. 1, C to E). O-lattice analysis explains how the network geometry, loop planes, and ledge-associated dislocation reactions arise from 3D lattice misfit and accommodate transformation strain. Together, these results connect point-defect transport, interfacial dislocation-network motion, migration anisotropy, and morphology evolution during microstructural processes governed by semicoherent interfaces.

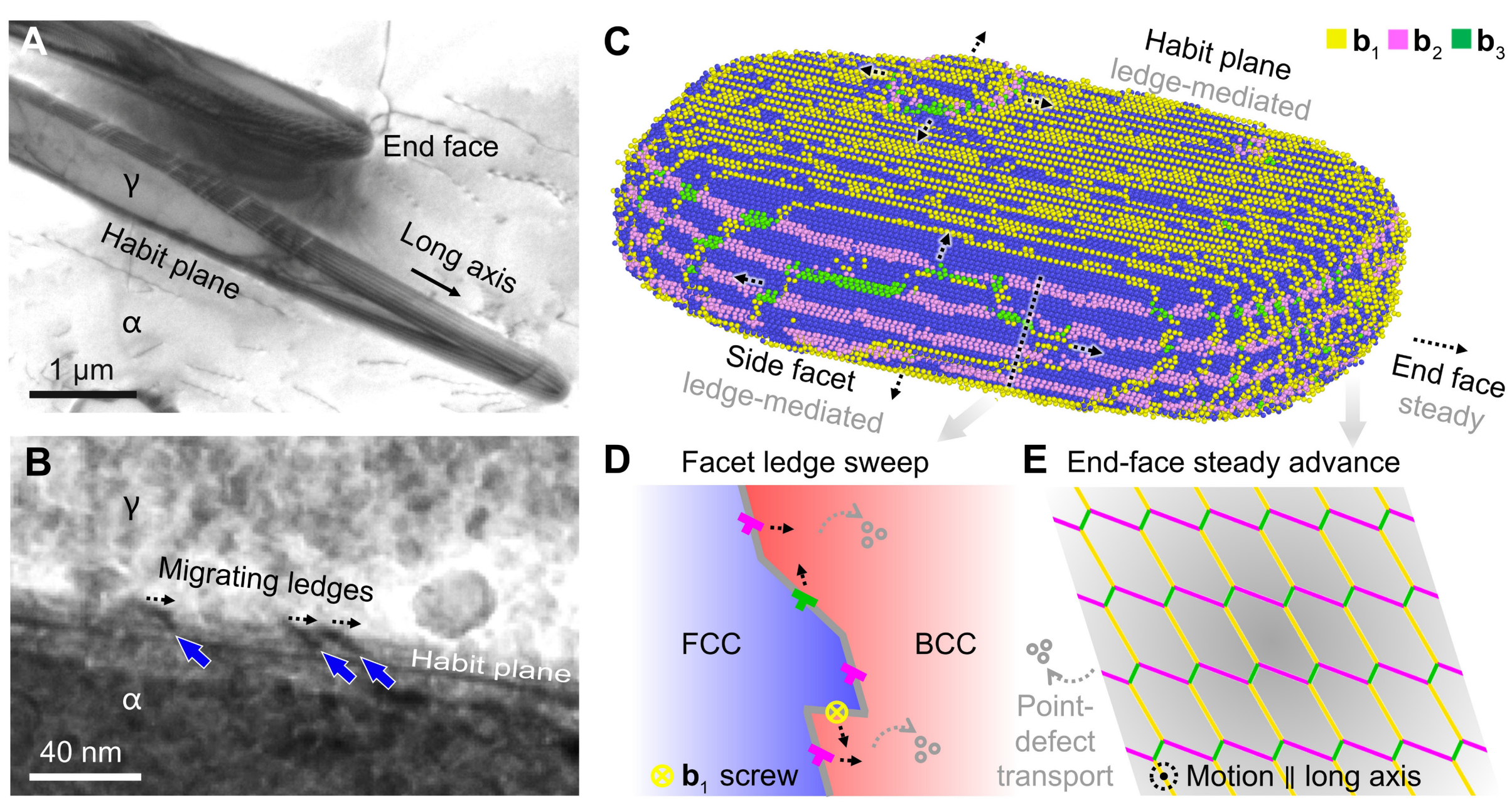

**Fig. 1. Overview of experimentally observed ledge-mediated interface advance and simulation-resolved dislocation-network dynamics.** (**A**) TEM image of a representative γ-austenite precipitate in an α-ferrite matrix, showing the broad habit plane, end face, precipitate long axis, and interfacial line contrasts associated with growth ledges and dislocation networks. (**B**) Representative in situ TEM frame from movie S1, acquired during in situ heating at 750 °C, showing growth ledges migrating on a γ/α habit plane. Blue arrows mark the ledges, and black dashed arrows indicate their lateral propagation along the habit plane approximately parallel to the precipitate long axis. (**C**) 3D PFC simulation snapshot showing the coupled migration modes around a lath-shaped FCC precipitate in a BCC matrix. BCC matrix atoms are omitted for clarity. Interfacial dislocations are colored according to their Burgers-vector content. The habit plane and side facet advance by ledge-mediated migration, whereas the end face advances steadily along the long axis. (**D**) Side-facet projection illustrating facet-ledge sweep through mixed glide–climb motion and local reactions of dislocation segments, assisted by point-defect transport, viewed along the long axis. (**E**) End-face projection showing steady advance through collective, non-conservative motion of the closed interfacial dislocation network along the long axis.

## Results

### Mixed-mode habit-plane migration revealed by in situ TEM

We first examine the migration of γ/α habit planes in freshly formed austenite precipitates within a ferrite matrix during in situ TEM heating. These precipitates are projected with their long axes approximately within the TEM foil plane, allowing both habit-plane sides to be observed (fig. S1). The habit plane advances in a mixed mode: discrete nanometer-high growth ledges propagate laterally along the precipitate long axis, while a concurrent normal advance occurs between resolvable ledge passages (Fig. 1B and movie S1). We then quantified these facet-migration kinetics in a near edge-on view. In a representative event at 745 °C, three nanometer-high growth ledges (Fig. 2, A, B, D, and E and movies S2 and S3) migrate laterally over tens of seconds, producing stepwise increments in habit-plane position. Frame-difference images further show that the interface continues to advance normally between successive ledge passages (Fig. 2, C and F). Quantitatively, the lateral ledge-propagation rate along the long axis is substantially higher than the average normal-advance rate ($\sim 10\ \mathrm{nm\ s^{-1}}$ versus $\sim 1.2\ \mathrm{nm\ s^{-1}}$), demonstrating strongly anisotropic habit-plane kinetics, which anchors the facet-migration mode later resolved in simulations.

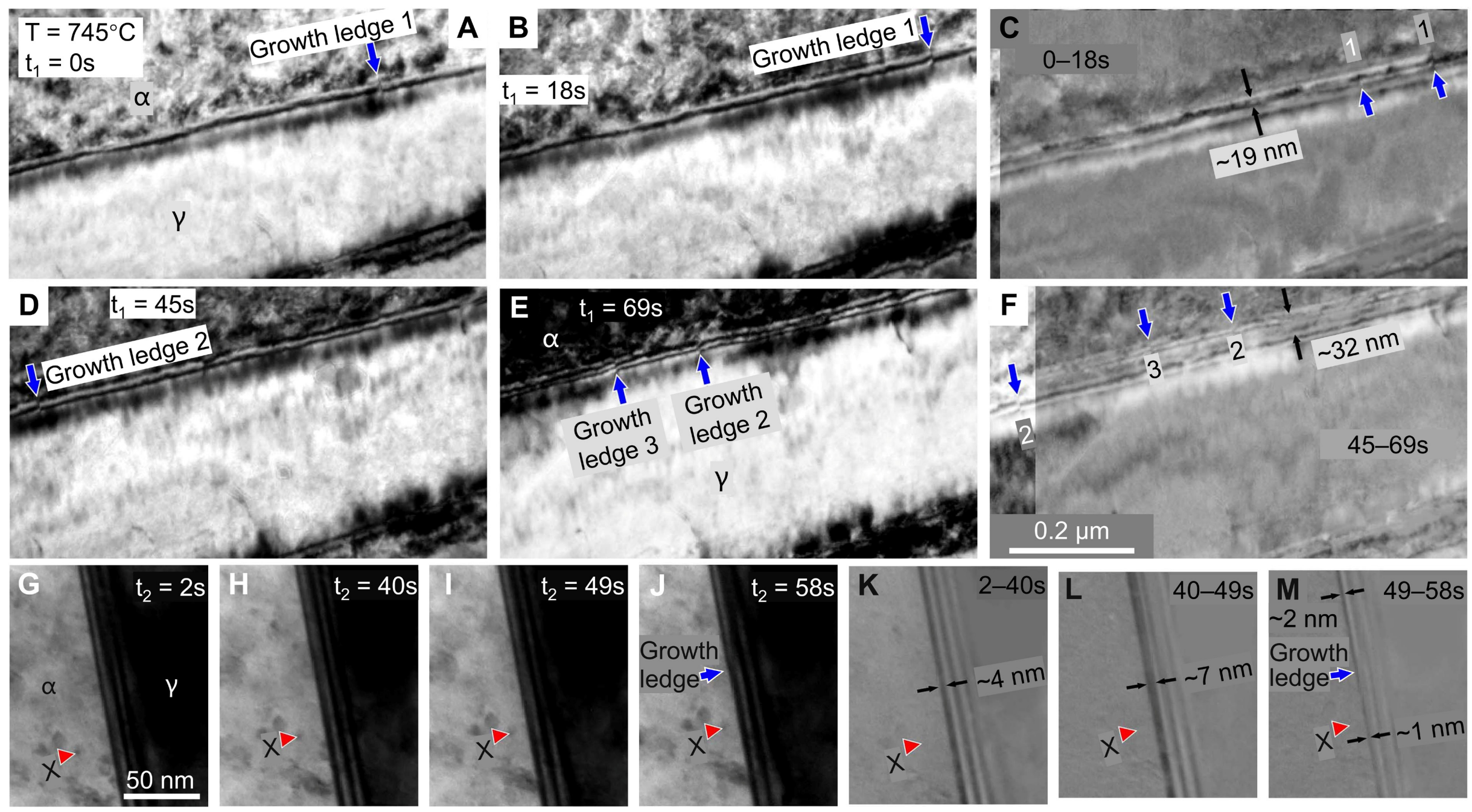

**Fig. 2. Mixed-mode habit-plane advance by lateral ledge propagation and concurrent normal advance during in situ TEM heating.** (**A**, **B**, **D**, and **E**) Sequential TEM frames from an in situ heating sequence at 745 °C, showing growth ledges propagating laterally along a γ/α habit plane. Blue arrows mark the ledges. (**C** and **F**) Frame-difference images for the 0–18 s and 45–69 s intervals, respectively, highlighting the net normal advance of the habit plane. Black arrow pairs mark the measured normal displacement, and numbered labels indicate the corresponding ledge positions before and after propagation. (**G** to **J**) Sequential higher-magnification TEM frames from a second in situ heating sequence at 750 °C, showing habit-plane motion and the passage of a growth ledge. The red arrowhead and X label mark a fiducial feature used for image-drift correction. (**K** to **M**) Frame-difference images for the 2–40 s, 40–49 s, and 49–63 s intervals, respectively. Black arrow pairs mark the projected normal displacements of the habit plane. In (M), the region swept by the growth ledge shows an additional displacement relative to the adjacent unswept region, providing an estimate of the projected ledge height.

At a second habit plane imaged at higher magnification (Fig. 2, G to J and movie S4), the two contributions to interface advance can be separated more clearly. Frame-difference images reveal an apparently ledge-free normal advance of the interface during intervals without a resolved ledge passage (Fig. 2, K and L). When a ledge is present, the swept region shows a larger normal advance than adjacent unswept regions over the same time window (Fig. 2M), allowing the ledge-associated increment to be distinguished from the background normal advance. The projected displacement difference (~1 nm) provides an estimate of the ledge step height, which is superposed on the concurrent normal advance. Together, these observations show that habit-plane migration proceeds through discrete ledge-mediated increments and a continuous normal advance.

**Three-dimensional dislocation-network dynamics during precipitate growth**

The in situ TEM observations establish ledge-mediated habit-plane migration, but the images at elevated temperatures cannot directly resolve the 3D motion of the underlying interfacial dislocation network. We therefore used 3D PFC simulations to track semicoherent precipitate growth in a model BCC-to-FCC transformation with a Kurdjumov–Sachs (K–S) orientation relationship. Starting from a spherical face-centered cubic (FCC) nucleus embedded in a body-centered cubic (BCC) matrix, the precipitate rapidly develops into a lath bounded by broad habit planes, narrow side facets, and curved end faces (Fig. 3 and movie S5). The long axis aligns with the parallel close-packed directions, $[0\bar{1}1]_f \parallel [1\bar{1}1]_b$. Both the habit planes and side facets remain semicoherent, accommodating misfit via periodic $\mathbf{b}_1$ and $\mathbf{b}_2$ interfacial dislocation arrays, respectively. On the end face, $\mathbf{b}_1$ and $\mathbf{b}_2$ segments intersect, with short $\mathbf{b}_3 = \mathbf{b}_1 + \mathbf{b}_2$ segments at some junctions. These key features, including lath shape, the major facets, and interfacial dislocation structures, are consistent with reports on lath austenite precipitates in duplex stainless steels (*12*), and lath Cr precipitate in Cu- (*20, 25*) and Ni-based alloys (*24*) (table S1), although the idealized lattice-parameter ratio in the PFC model introduces slight deviations in long-axis and facet orientations and suppresses additional coarse-spaced side-facet dislocations (*19, 23, 24*) (table S1).

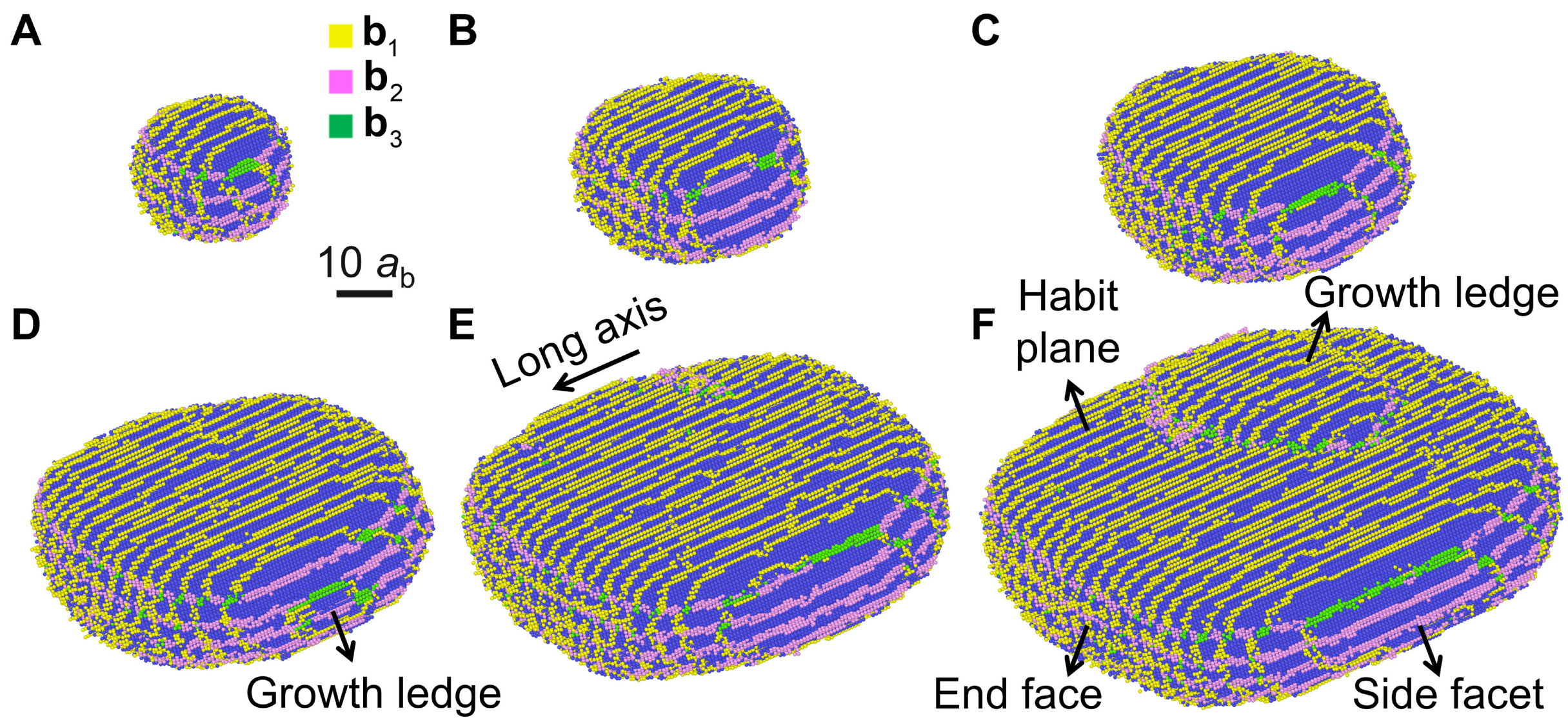

**Fig. 3. Growth process of the FCC precipitate in the BCC matrix.** K–S orientation relationship variant is $[0\bar{1}1]_f \parallel [1\bar{1}1]_b$, $(111)_f \parallel (011)_b$. Atoms belonging to the FCC phase are shown in blue. Atoms at the dislocation cores with different Burgers vectors are colored distinctly ($\mathbf{b}_1 = [0\bar{1}1]_f/2 \mid [1\bar{1}1]_b/2$, $\mathbf{b}_2 = [10\bar{1}]_f/2 \mid [11\bar{1}]_b/2$, and $\mathbf{b}_3 = [1\bar{1}0]_f/2 \mid [100]_b$). For visual clarity, atoms in the surrounding BCC matrix are removed.

The lath morphology arises from strongly mode-dependent interface migrations. Despite carrying a dense dislocation network, the end face advances rapidly and steadily along the long axis. In contrast, the side facet and habit plane migrate intermittently through a ledge mechanism: a semicoherent ledge loop nucleates on either facet and propagates laterally (Fig. 3, D to F). Accompanied by their associated dislocations, ledge propagation is also highly anisotropic—steady along the long axis but kink-mediated in the transverse direction. Each ledge sweep advances the corresponding facet by one ledge height, mainly contributing to lath thickening. When a ledge reaches a facet junction or end face, its riser and associated dislocation(s) are incorporated into the adjoining interface. The PFC-predicted ledges are consistent with the growth ledges reported on semicoherent FCC/BCC facets in multiple alloy systems, with a typical height of 1 nm (*18, 19, 23, 26*).

Compared with the side facet, ledge nucleation on the habit plane is more intermittent, consistent with its greater stability: its screw-character $\mathbf{b}_1$ dislocation arrays make it energetically more stable than the side facet (*45*) and therefore more resistant to ledge nucleation. Once nucleated, however, habit-plane ledges do not exhibit the same pronounced kink-controlled anisotropy as side-facet ledges. This difference probably reflects the risers' crystallographic character: side-facet ledges have habit-plane-like risers with mainly screw-character dislocations that resist kink formation, whereas habit-plane ledges have side-facet-like risers that make kink generation easier.

The dislocations on the habit planes, side facets, and end faces remain connected into closed loops encircling the precipitate, consistent with previous interpretations of semicoherent precipitate interfaces (*9, 25*). Time-resolved tracking reveals that these loops expand within well-defined loop planes that serve as pathways for dislocation motion and reorganization during precipitate growth. Specifically, $\mathbf{b}_1$ dislocations on the habit planes and end faces join into loops parallel to the side facet, whereas $\mathbf{b}_2$ dislocations on the side facets and end faces reside in loops parallel to the habit plane. These loop planes are generally not typical slip planes because the semicoherent facets usually have irrational orientations when they contain one set of dislocations.

As a result, loop expansion often proceeds by mixed glide–climb motion of dislocations and requires atomic diffusion. For example, the Burgers vector of $\mathbf{b}_2$ is inclined by ~22° relative to its loop plane, implying a climb component when the dislocation moves along the loop plane as the loop expands. Consistently, fig. S2 shows that during the steady migration of the end face, $\mathbf{b}_2$ dislocations undergo non-conservative motion with a negative climb component, as expected when BCC transforms into the denser FCC structure. Overall, our simulation provides a 3D, time-resolved view that links non-conservative interfacial dislocation motion to facet/end-face migration modes and, ultimately, to Widmanstätten-type lath morphology widely observed in structural alloys.

**Crystallographic basis of the dislocation-network geometry and motion**

To interpret the interfacial defect structures that emerge during the simulated precipitate growth, we analyzed the semicoherent FCC/BCC interfaces using O-lattice theory (*44*), which assumes that the mismatch between the two lattices is fully accommodated by interfacial dislocations. In the long-axis projection (Fig. 4A), the edge-on O-lines (centers of coherent patches) and O-cell walls (loci of maximum misfit) reflect the periodicity of the fit–misfit pattern. The habit plane and side facet are defined by the planes that intersect one set of O-lines/O-cell walls and contain, at the intersections with O-cell walls, one set of dislocations: the $\mathbf{b}_1$ array and the $\mathbf{b}_2$ array, respectively. A network consisting of $\mathbf{b}_1$, $\mathbf{b}_2$, and $\mathbf{b}_3$ dislocation segments forms on the end face, which intersects multiple sets of O-cell walls. The close agreement among the dislocation structure based on the misfit distribution (Fig. 4A), the dislocation cores resolved in the 3D PFC simulations (Fig. 4, B and C), and the experimental high-resolution TEM observations (e.g., Figs. 3 and 6C by Furuhara et al. (*23*)) indicates that the dislocation network preserved during precipitate growth is largely prescribed by the misfit distribution. Extending the O-cell-wall construction to the full 3D precipitate (Fig. 5) further predicts a closed interfacial network, in which segments on different interfaces connect across edges.

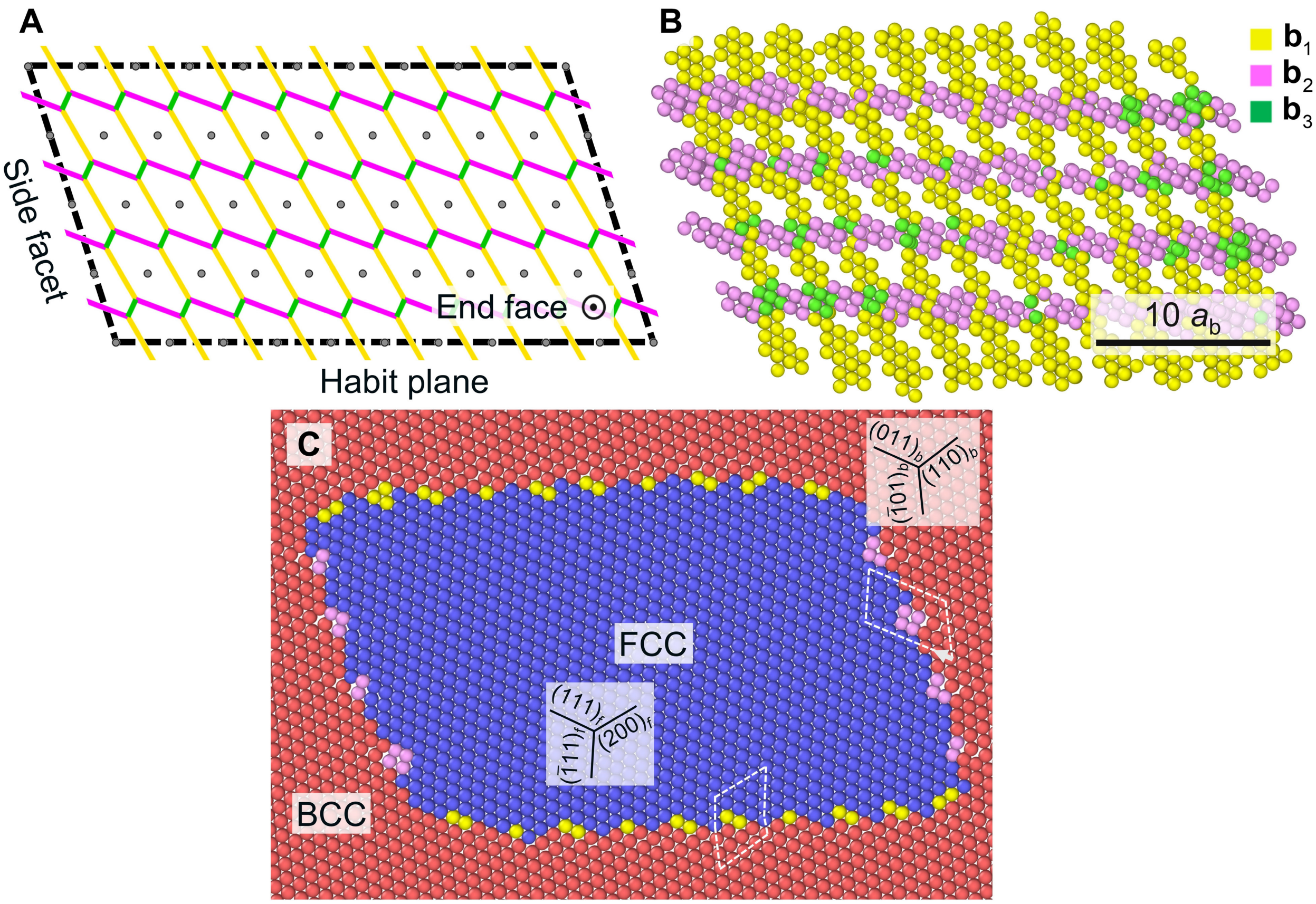

**Fig. 4. O-lattice prediction and PFC simulations of interfacial dislocations surrounding the precipitate.** (**A**) O-lattice construction viewed from the long axis: gray dots mark edge-on O-lines and colored segments indicate edge-on O-cell walls associated with different Burgers vectors; dashed lines outline the traces of the habit plane and side facet, while the end-face region is nearly normal to the long axis. (**B**) A PFC snapshot showing the corresponding interfacial dislocation

network. All the atoms inside the crystals are removed. (**C**) The cross-sectional view of the precipitate. Atoms in FCC and BCC are colored blue and red, respectively. In (C), the $\mathbf{b}_1$ dislocation core is not visible in a 2D Burgers circuit because of no in-plane Burgers-vector component.

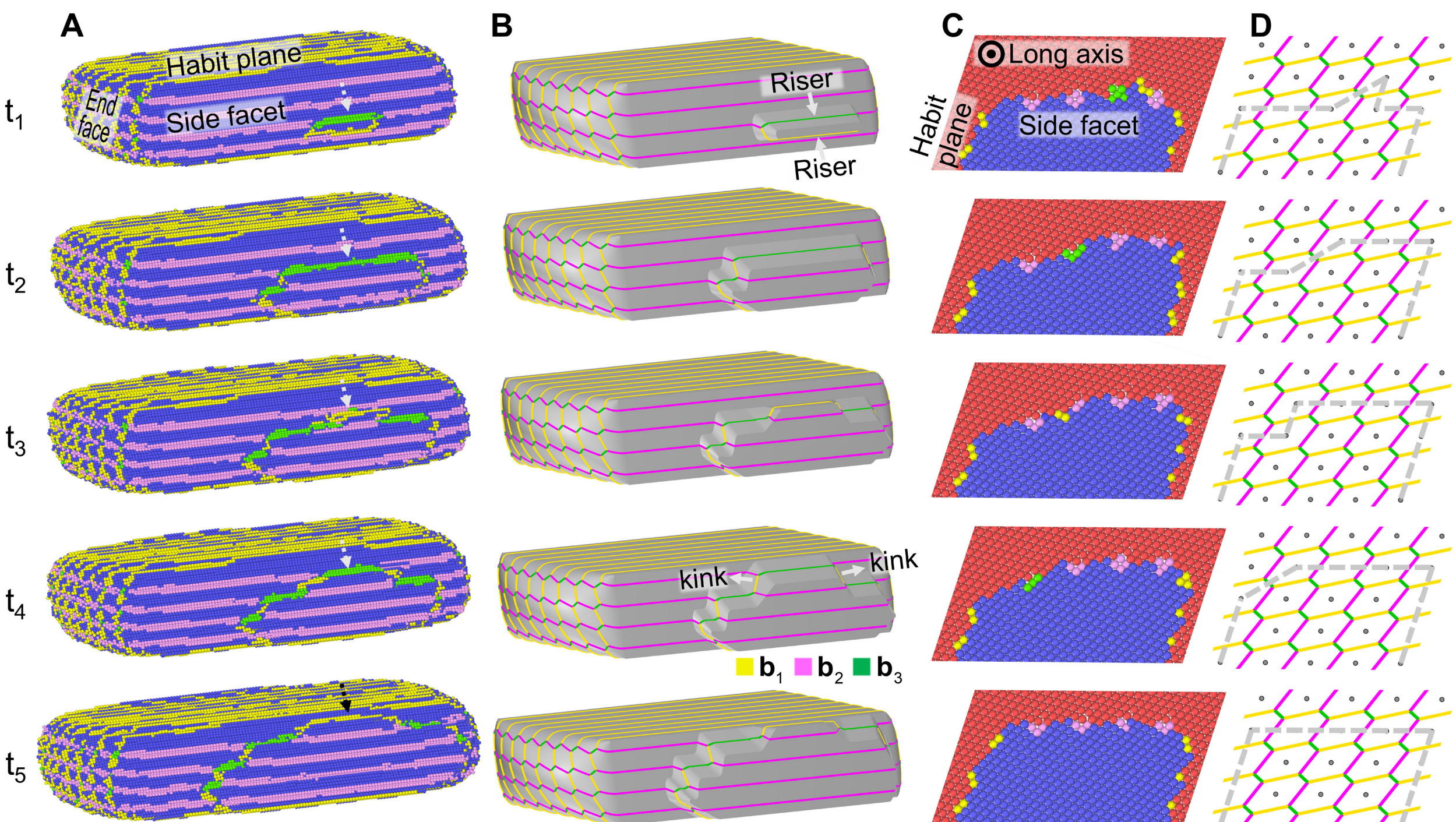

**Fig. 5. Growth-ledge nucleation on the side facet ($t_1$), kink-mediated riser propagation ($t_2$ to $t_4$), and riser reincorporation into adjoining facets ($t_5$).** (**A**) PFC snapshots. (**B**) Corresponding 3D dislocation network obtained from O-lattice analysis, taking the interface geometry from the simulated precipitate. (**C**) Cross-sectional views taken through the planes perpendicular to the long axis, whose positions are indicated by arrows in (A). (**D**) Corresponding O-lattice constructions viewed along the invariant-line direction. Atoms in FCC and BCC are colored blue and red, respectively. In (A to C), dislocations are colored according to their Burgers vectors. In (D), gray dots denote edge-on O-lines, colored segments denote edge-on O-cell walls, and gray dashed lines trace the interface and ledge geometry.

We use the side facet as an example to resolve how growth ledges nucleate and propagate with local reorganization of the interfacial dislocation network. The instantaneous PFC interface geometry (Fig. 5, A and C, and movie S6) was used as input for the O-lattice construction (Fig. 5,

B and D) to illustrate the dislocation reactions observed in the PFC simulations. As seen in Fig. 5, A to D, $t_1$, ledge nucleation initiates by splitting a $\mathbf{b}_2$ dislocation on the side facet into $-\mathbf{b}_1$ and $\mathbf{b}_3$ segments, located at two ledge risers with distinct orientations. The transverse advance of the riser proceeds through a kink-mediated process. Specifically, kink nucleation on the riser decorated with $\mathbf{b}_3$ is associated with local decomposition of a $\mathbf{b}_3$ segment into $\mathbf{b}_1$ and $\mathbf{b}_2$ segments, with the $\mathbf{b}_2$ segment extending onto the ledge terrace (Fig. 5, A to D, $t_3$). The newly formed $\mathbf{b}_1$ segment then combines with a neighboring $\mathbf{b}_2$ line on the original terrace, thereby restoring the $\mathbf{b}_3$ segment on the advanced ledge riser (Fig. 5, A to D, $t_4$). This reaction sequence forms a pair of kink planes decorated by short $\mathbf{b}_1$ segments (Fig. 5, A and B, $t_4$), which then sweep rapidly along the long-axis direction. As the riser merges back into the facet, the $\mathbf{b}_3$ segment further decomposes into $\mathbf{b}_1$ and $\mathbf{b}_2$, restoring the dominant $\mathbf{b}_1$ and $\mathbf{b}_2$ dislocation lines on the facets (Fig. 5, A to D, $t_5$).

The PFC simulation (Fig. 5, A and C) reveals that ledge nucleation and propagation during semicoherent interface migration are mediated by dislocation reactions, which repeatedly rearrange the interfacial dislocation segments on the riser and terrace of a ledge. These dislocation segments retain their misfit-accommodating role across the terraces, ledges, and kinks, with their locations as the intersections of O-cell walls and corresponding portions of the interface.

This O-lattice-based interpretation is not limited to the idealized lattice-parameter ratio used in the structural PFC model. For a given crystallographic system, the same construction can be carried out using its actual lattice parameters, allowing the corresponding growth-ledge geometry and dislocation arrangement to be determined, as demonstrated in our prior study (*29*) and in additional examples in the Supplementary Information.

## Discussion

Two distinct modes of interface migration operate during semicoherent lath growth. Following Cahn's description (*46*), an interface may advance either by uniform normal motion or by lateral step propagation, depending on the driving force relative to the resistance arising from the misfit-associated energy variations with interface position. In the present system, this resistance varies substantially across different interface regions surrounding the precipitate, giving rise to distinct migration modes that account for the pronounced lath morphology. Our results further reveal the fine details of how the migration of habit planes, side facets, and end faces is coupled through the reorganization of a closed network of misfit dislocations.

The variation in interfacial energy arises not only from the continuous anisotropic misfit distribution but also from the discrete dislocation structures that accommodate misfit, whose geometry is dictated by the intersections between the interface and the O-cell walls (Fig. 5). Because the O-cell walls extend along the invariant line, the dislocation configuration and the associated fit–misfit pattern on the end face are largely preserved as the end face advances along the invariant line, that is, the lath long axis. Hence, a much lower lattice resistance is expected for end-face advance, where different dislocation segments can move collectively and non-conservatively with the migrating interface without major reconstruction of the network. This explains the rapid and steady lengthening of the lath through end-face migration.

In contrast, the habit planes, side facets, and ledge risers are constrained by the discrete positions of O-lines (Fig. 5). Any normal advance away from these low-misfit configurations generates excess misfit strain that cannot be accommodated by the existing interfacial dislocations, giving rise to a lattice resistance to uniform motion. This resistance can, in principle, be surmounted in two ways. At a sufficiently high driving force, the interface may undergo uniform, shear-coupled motion, wherein relative lattice displacement enables the interface to retain a low-misfit structure during normal advance. At lower driving forces, the facet advances instead by lateral ledge migration, in which semicoherent growth ledges connect neighboring low-misfit terrace positions. In diffusional transformations such as precipitate growth during aging, where point-defect transport enables non-conservative dislocation motion, the broad facets follow the latter pathway: growth ledges nucleate and propagate laterally. The nucleation of a ledge involves the generation of dislocations in the riser, accompanied by various local dislocation reactions. Meanwhile, the characteristic terrace features are sustained and transferred to the advanced low-misfit position through the decomposition and recombination of dislocation segments. These intricate dislocation reactions effectively accommodate the misfit strain while limiting both the rate of ledge nucleation and the propagation of kink-mediated ledges. Consequently, lath thickening and widening proceed much more slowly than lengthening, which is not limited by such reactions. In situ TEM further suggests an intermediate behavior on real habit planes, where discrete ledge passage is superposed on an apparently ledge-free normal advance at the experimental spatiotemporal resolution, likely facilitated by surface strain relief in thin foils.

Taken together, these results suggest a hierarchical picture of how semicoherent interfaces advance during 3D precipitate growth (Fig. 1): point-defect transport (point-defect level) regulates interfacial dislocation reactions (line-defect level, Fig. 1, D and E); those reactions determine the local interface migration modes and misfit accommodation (planar-defect level, Fig. 1, D and E), and the accumulated local modes drive anisotropic 3D growth and morphology evolution (bulk level, Fig. 1C). The O-lattice/O-cell construction shows that the same connected dislocation network that mediates interface migration also preserves misfit accommodation and transformation-strain compatibility as the precipitate grows.

This work therefore answers how habit planes, side facets, and end faces migrate together during growth mediated by semicoherent interfaces: they do not advance as independent interface regions, but as parts of a coupled, evolving interfacial dislocation network. By linking point-defect transport, dislocation-network motion, ledge propagation, interface-migration anisotropy, and morphology evolution, this framework connects crystallographic descriptions of semicoherent interfaces to their kinetic behavior. More broadly, this framework provides a transferable basis for predicting defect-mediated microstructural evolution in processes governed by moving semicoherent interfaces, with potential relevance to other crystallographically constrained, lattice-mismatched systems.

## Materials and Methods

### In situ TEM experiments

A commercial duplex stainless steel with nominal composition Fe–24.9Cr–7.0Ni–3.1Mo (wt%) was used. For in situ observations of freshly nucleated austenite, 10×10×10 $mm^3$ blocks were encapsulated in evacuated silica tubes and solution-treated at 1300 °C for 30 min, followed by water quenching. TEM foils were prepared from 0.5-mm-thick slices by mechanical thinning and twin-jet electropolishing (Struers Tenupol-3) using 8 vol% perchloric acid in ethanol at 20 V and −30 °C. Immediately before heating experiments, foils were plasma cleaned (Solarus, Gatan; 1 min) to remove hydrocarbon contamination. In situ heating was conducted in a Philips CM20FEG microscope operated at 200 kV using a Gatan double-tilt heating holder. Samples were first heated to 700 °C at the maximum heating rate (~50 °C/min), then heated in 5 °C increments until interface migration or new austenite growth was observed (typically 720–850 °C). Videos were recorded using a Gatan Orius side-entry CCD camera at 15–24 frames per second, and key frames were extracted for subsequent analysis.

**Phase-field crystal model**

We employed the structural PFC model developed by Greenwood et al. (*47, 48*) to simulate phase transformations in an FCC/BCC system. The state variable is the dimensionless atomic density field $n(\mathbf{r})$. The free-energy functional takes the form (*48*):

$$F = \int d\mathbf{r}\big(n(\mathbf{r})^2/2 - \eta n(\mathbf{r})^3/6 + \chi n(\mathbf{r})^4/12\big) - \int d\mathbf{r} \int d\mathbf{r}'\, n(\mathbf{r}) C_2(|\mathbf{r} - \mathbf{r}'|) n(\mathbf{r}')/2, \quad (1)$$

where the first term is a local Landau-type polynomial expansion around a uniform reference state, and the second term introduces crystallographic order through the two-body correlation kernel $C_2(|\mathbf{r}-\mathbf{r}'|)$. Following Greenwood et al., $C_2$ was constructed in reciprocal space as an envelope of Gaussian peaks (*47, 48*):

$$C_2^i(k) = \exp\left(-\sigma^2/\sigma_{\mathrm{M}i}^2\right)\exp\left(-(k-k_i)^2/2\alpha_i^2\right), \qquad (2)$$

where $k_i$, $\alpha_i$, and $\sigma_{\mathrm{M}i}$ control the peak position, width, and the dependence on the effective-temperature parameter $\sigma$, respectively. To stabilize the FCC phase with a dimensionless lattice parameter $a_f = \sqrt{3/2}$, two peaks were used at $k_1 = 2\sqrt{3}\pi/a_f$ and $k_2 = 4\pi/a_f$, corresponding to $\{111\}_f$ and $\{200\}_f$ plane families, respectively. Under this parameterization, the BCC lattice parameter satisfies $a_b = 2\sqrt{2}\pi/k_1 = 1$ because the first peak corresponds to $\{110\}_b$. The bulk free-energy densities of the FCC and BCC phases were evaluated as a function of the effective-temperature parameter $\sigma$ following the method in a prior study (*47*) (fig. S3), which identifies the phase-stability crossover used to select transformation conditions.

The density field $n(\mathbf{r})$ evolved according to conserved dynamics (*41*):

$$\partial n(\mathbf{r})/\partial t = \nabla^2[\delta F/\delta n(\mathbf{r})] + \xi, \qquad (3)$$

where $\xi$ represents conserved Gaussian colored noise subject to a high-frequency cutoff, which filters out unphysical fluctuations with wavelengths shorter than the atomic spacing (*49*). This equation conserves the spatial integral of $n(\mathbf{r})$ and enables evolutions on diffusive timescales within the PFC framework. Time integration was performed in reciprocal space using a semi-implicit Fourier-spectral scheme. The dimensionless grid spacing and time step were $\mathrm{d}x = 0.125$ and $\mathrm{d}t = 0.001$, respectively. Simulations were carried out in a fixed box of 1152×648×432 grid points under fully periodic boundary conditions. We emphasize that the structural PFC simulations are used here to elucidate the coupled dislocation–ledge mechanism and the geometry-governed interface dynamics, rather than to quantitatively calibrate the dimensionless PFC time to experimental absolute timescales.

**Initialization and analysis workflow**

The initial configuration consisted of a spherical FCC precipitate embedded in a BCC matrix, with a dimensionless radius of 10 to suppress capillarity-driven shrinkage. The initial orientation relationship was set to the K–S orientation relationship with the variant: $[0\bar{1}1]_f \parallel [1\bar{1}1]_b$ and $(111)_f \parallel (011)_b$. By adopting idealized lattice parameters ($a_f = \sqrt{3/2}a_b$), $[0\bar{1}1]_f/2$ and $[1\bar{1}1]_b/2$ have identical lengths, and they define the direction of the invariant-line. In addition, $(111)_f \parallel$ $(011)_b$ define the direction of the invariant-line in reciprocal space. Therefore, three sets of O-lines are solvable for this special case, which simplifies the dislocation structures while retaining the key process of the dislocation-mediated ledge mechanism of growth.

After simulation, atomic positions were extracted from the local maxima (density peaks) of $n(\mathbf{r})$. Local crystal structures (FCC/BCC/other) were identified using polyhedral template matching (*50*) as implemented in OVITO (*51*). Interfacial dislocations were characterized by their spatial distribution and Burgers vectors, determined from the Nye tensor field evaluated on the extracted atomic configuration. The Nye tensor was computed and decomposed using singular value decomposition to obtain the dominant Burgers-vector content (*52*).

**Crystallographic calculation of interfacial dislocations**

Interfacial dislocation structures were determined corresponding to the locations of maximal misfit. The O-lattice theory (*44*) provides a general tool for analyzing misfit distribution on any interface. For the K–S orientation relationship considered here, using the special lattice parameter ratio, the misfit at any interface can be accommodated by interfacial dislocations whose Burgers vectors lie within the parallel planes $(111)_f \parallel (011)_b$, namely, $\mathbf{b}_1 = [0\bar{1}1]_f/2 \mid [1\bar{1}1]_b/2$, $\mathbf{b}_2 = [10\bar{1}]_f/2 \mid [11\bar{1}]_b/2$, and $\mathbf{b}_3 = [1\bar{1}0]_f/2 \mid [100]_b$, where the symbol "|" denotes corresponding vectors across the transformation. The position of an O-lattice element (position of zero misfit) is defined by a vector, $\mathbf{x}^{\mathrm{o}}$, solved from: $\mathbf{T}\mathbf{x}^{\mathrm{o}} = \sum_i k_i \mathbf{b}_{i\alpha}$ (*44*), where $k_i$ are integers, $\sum_i k_i \mathbf{b}_{i\alpha}$ indicates any translational vectors in the $\alpha$ phase. $\mathbf{T} = \mathbf{I} - \mathbf{A}^{-1}$ is the displacement matrix that links a position $\mathbf{x}_\gamma$ with its associated displacement $\mathbf{x}_\gamma - \mathbf{x}_\alpha$, which are related by $\mathbf{x}_\gamma = \mathbf{A}\mathbf{x}_\alpha$. Here, $\mathbf{A}$ represents the deformation matrix transforming the $\alpha$ lattice to the $\gamma$ lattice. A set of O-cell walls between O-elements corresponds to positions of maximal misfit and is defined by a reciprocal vector, $\mathbf{c}^{\mathrm{o}}$, related to a Burgers vector, $\mathbf{b}_{i\alpha}$, by $\mathbf{c}^{\mathrm{o}} = \mathbf{T}^{\mathrm{T}}\, \mathbf{b}_{i\alpha}/|\mathbf{b}_{i\alpha}|^2$ (*53*). The dislocations are defined by the intersections between the O-cell walls and an interface, and the Burgers vector of the dislocations is $\mathbf{b}_{i\alpha}$ associated with the particular set of O-cell walls. The interfacial dislocations determined based on the O-lattice theory were employed to compare with PFC-predicted defect structures and experimental observations.

## Acknowledgments

J.-Y.Z. thanks Prof. Nikolas Provatas at McGill University and Prof. Nana Ofori-Opoku at McMaster University for instruction on the use of structural PFC codes, and Dr. Rui-Xun Xie at Tsinghua University for technical assistance in setting up the server used for PFC simulations.

**Funding:**

Japan Society for the Promotion of Science (JSPS) KAKENHI grant JP23K20037 (J.-Y.Z. and S.O.).

Ministry of Education, Culture, Sports, Science and Technology (MEXT), Japan, program grant JPMXP1122684766 (S.O.).

National Natural Science Foundation of China grant 51871131 (W.-Z.Z.).

National Key Research and Development Program of China grant 2016YFB0701304 (W.-Z.Z.).

China Scholarship Council scholarship 201906210309 (J.-Y.Z.).

China Scholarship Council scholarship 201506210313 (J.D.).

**Author contributions:**

Conceptualization: J.-Y.Z. and W.-Z.Z.

Methodology: J.-Y.Z., L.Y., and F.M.

Investigation: J.-Y.Z. and J.D.

Visualization: J.-Y.Z. and J.D.

Supervision: S.O. and W.-Z.Z.

Resources: S.O. and W.-Z.Z.

Writing—original draft: J.-Y.Z. and S.O.

Writing—review & editing: J.-Y.Z., L.Y., F.M., S.O., and W.-Z.Z.

# Supplementary Materials for

## Semicoherent interfaces advance through closed dislocation networks that cannot glide alone

Jin-Yu Zhang *et al.*

*Corresponding authors. Jin-Yu Zhang, jinyu.zhang@tsme.me.es.osaka-u.ac.jp; Shigenobu Ogata, ogata@me.es.osaka-u.ac.jp; Wen-Zheng Zhang, zhangwz@tsinghua.edu.cn

**This PDF file includes:**

## Supplementary Text

Using experimental lattice parameters, we calculated the orientation relationship and habit plane structures by the O-line model (*54*) implemented in PTCLab (*55*). The lattice correspondence between the BCC and FCC phases for the K–S variant considered in this study is:

$$\mathbf{C}_{b\rightarrow f} = \begin{bmatrix} 0.5 & 0.5 & 0 \\ -0.5 & 0.5 & 0 \\ 0 & 0 & 1 \end{bmatrix}.$$

The Burgers vector on the habit plane, i.e., $\mathbf{b}_1$, is taken as the input. Consistent with the experimentally observed orientation relationship in Cu–Cr alloy, the angle between the corresponding Burgers vectors $\mathbf{b}_1$ in the two phases, i.e., $[0\bar{1}1]_f/2$ and $[1\bar{1}1]_b/2$, is set as 0.5°.

When the experimental lattice parameters are used, $[0\bar{1}1]_f/2$ and $[1\bar{1}1]_b/2$ generally have different magnitudes, and the direct/reciprocal invariant lines can deviate from the close-packed directions. As a result, O-line solutions may not exist for interface planes away from the habit plane. In such cases, we employ a generalized O-element approach developed in our prior work (*29*), which enables the determination of coherent regions and interfacial dislocation structures on different interfaces, including the side facets, end faces, and growth ledges.

Fig. S4 presents our geometric calculations of the faceting and interfacial defect structures of a Cr precipitate in a Cu matrix for comparison with previously reported experimental observations. In the previous study, four facets (F1–F4) were identified from TEM observations (Fig. 1A of Luo et al. (*25*)). Periodic ledges with different spacings were reported on facets F2 and F3 (schematically illustrated in Fig. 2B of Luo et al. (*25*)), whereas two families of dislocation arrays with fine and coarse spacings were observed on the curved facet F4. In contrast, dislocations on F1 were not reported, likely because their spacing is too small to produce resolvable contrast under the imaging conditions.

These facet morphologies and defect configurations are reproduced by our generalized O-element construction, in which neighboring generalized O-elements are connected to form the facets and growth ledges (Fig. S4). The quantitative comparisons in Fig. S4 and Table S2 demonstrate that the model captures key crystallographic and geometric features, including facet orientations, ledge spacings/heights on F2 and F3, and dislocation spacings on F4.

We further demonstrate the applicability of our geometric framework to a different system, namely HCP/BCC interfaces. The required inputs for the O-line model and the generalized O-element approach follow our previous studies (*56, 57*). In the Ti–7.26 wt%Cr alloy, growth ledges with a height of ~10 nm are observed on the habit plane (Fig. S5A, enlarged view), with faint contrast associated with the riser and periodic contrasts on the terrace. These ledge characteristics—including the terrace width, ledge height, and riser orientation—are reproduced by the generalized O-element/O-cell-wall construction (Fig. S5B). In the construction, the terrace (habit plane) contains a periodic array of dislocations, whereas the ledge riser is accommodated by multiple fine-spaced dislocations together with an additional coarse-spaced dislocation. The ledge risers can therefore be regarded as micro side facets; as they accumulate, they form the side facet of the precipitate.

Additionally, less frequently observed facet configurations can also arise. For example, the alternative side facet (labeled “side facet B” in Fig. S6B) is reproduced by choosing a different coarse-spaced dislocation array on the side facet, resulting in a distinct facet orientation compared with “side facet A”.

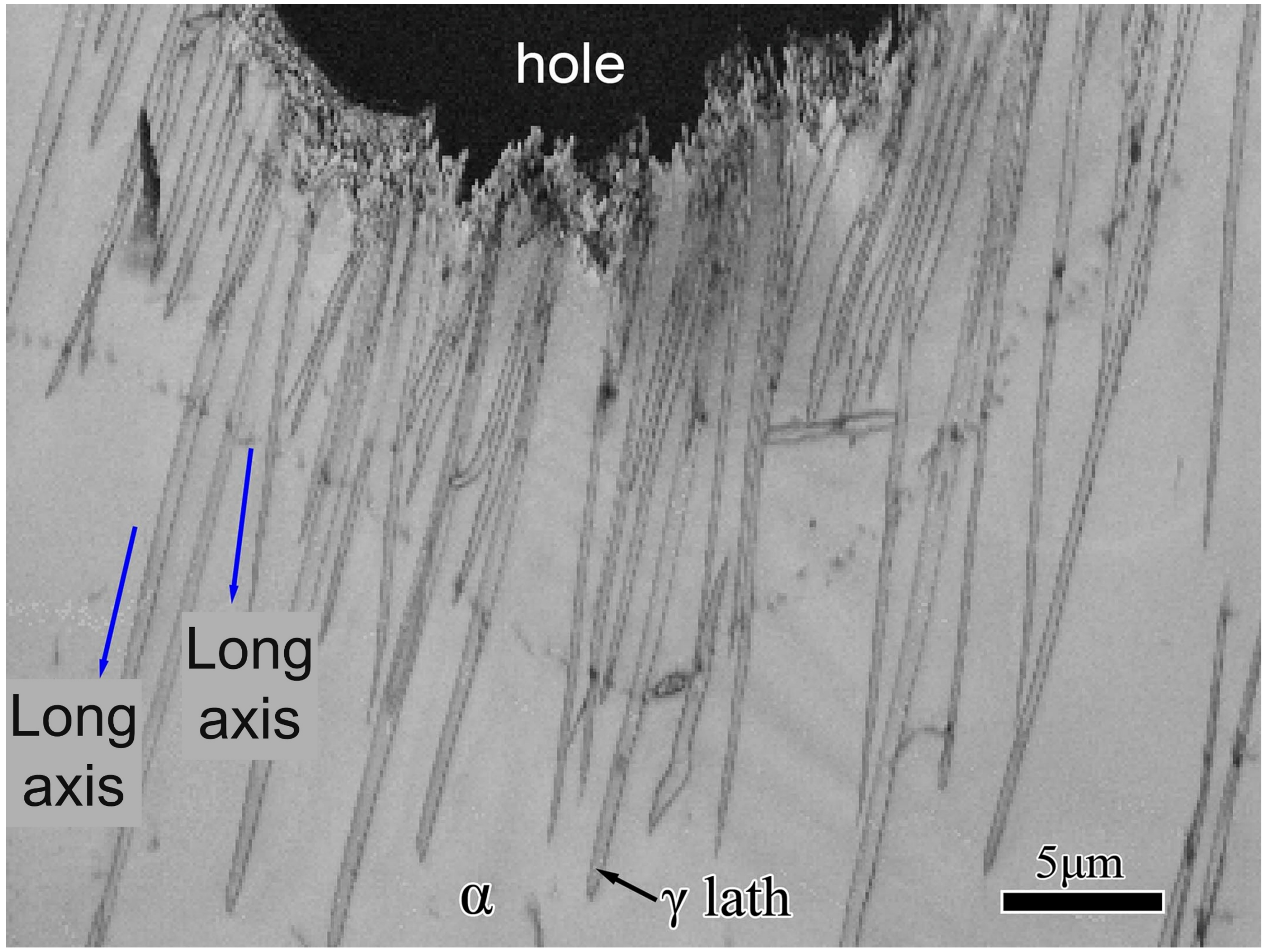

**Fig. S1. Post-mortem overview of newly formed austenite laths after in situ TEM heating.** Electron backscatter diffraction (EBSD) band-contrast map acquired near the perforation edge of the electropolished foil after the in situ heating experiment, showing newly formed austenite laths embedded in the ferrite matrix. Blue arrows indicate representative long-axis directions, revealing that the laths’ long axes are predominantly aligned within the foil plane.

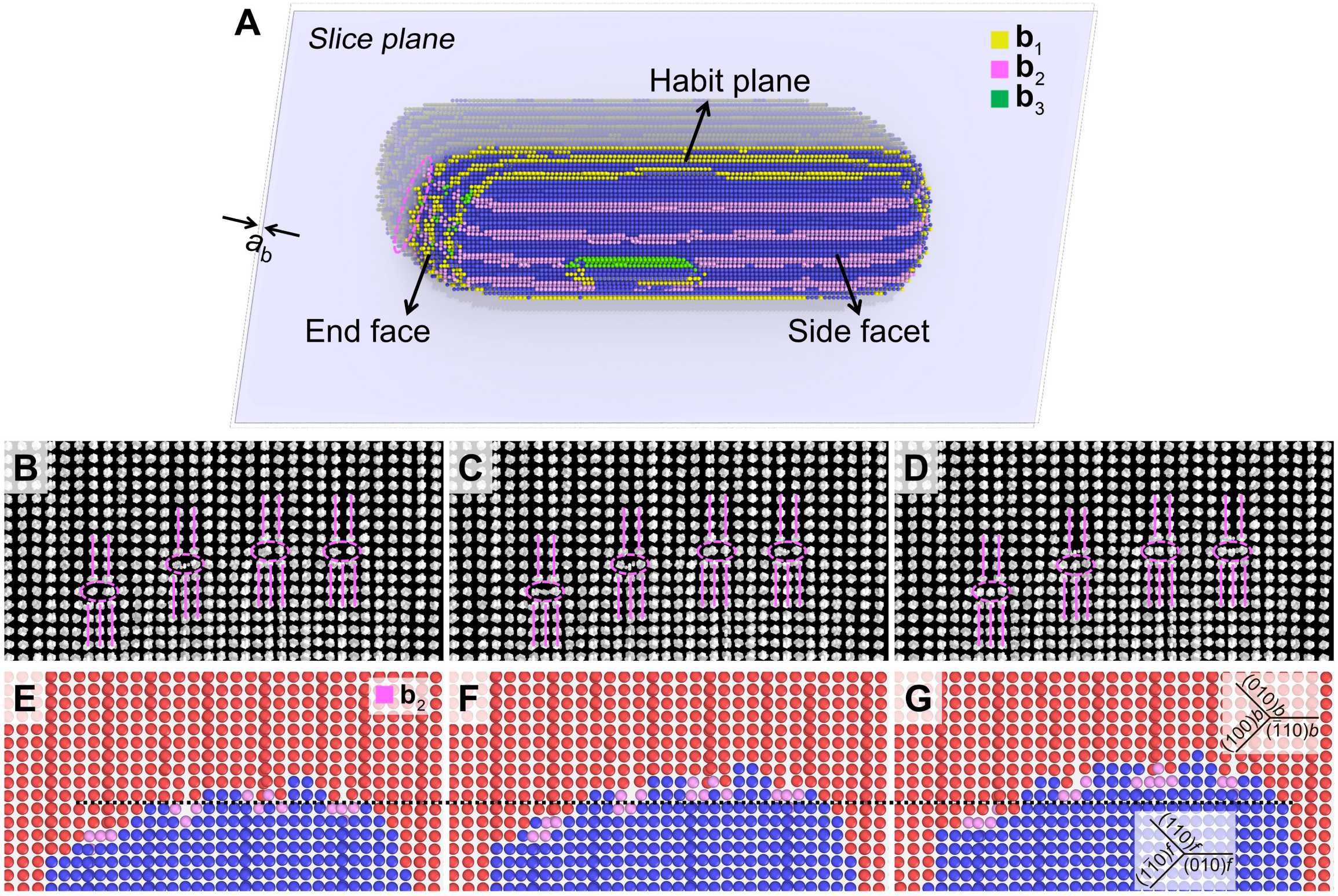

**Fig. S2. Climb of $\mathbf{b}_2$ dislocations during end-face migration.** (**A**) Schematic showing the slice geometry used for the local analysis. The precipitate is sliced by a slab parallel to the side facet and located in the end-face region (indicated by a dashed ellipse) between two neighboring $\mathbf{b}_1$ dislocation loops, so that only the $\mathbf{b}_2$ dislocation arrays are intersected. The sliced configurations, shown in (B to D), are viewed along $[001]_f \| [001]_b$. (**B** to **D**), Evolution of the thresholded PFC atomic density field ($n(\mathbf{r})>1.5$) during end-face migration. The solid lines mark the extra half-planes associated with the $\mathbf{b}_2$ dislocations, and the dashed ellipses indicate the core regions where climb occurs through an increase in local atom number. During this process, each $\mathbf{b}_2$ dislocation climbs upward by one atomic plane. (**E** to **G**), Corresponding atomic configurations. The $\mathbf{b}_2$ dislocation cores, FCC, and BCC atoms are shown in magenta, blue, and red, respectively. The horizontal dashed guide line across (E to G) assists in showing the upward climb of the dislocations by one atomic layer.

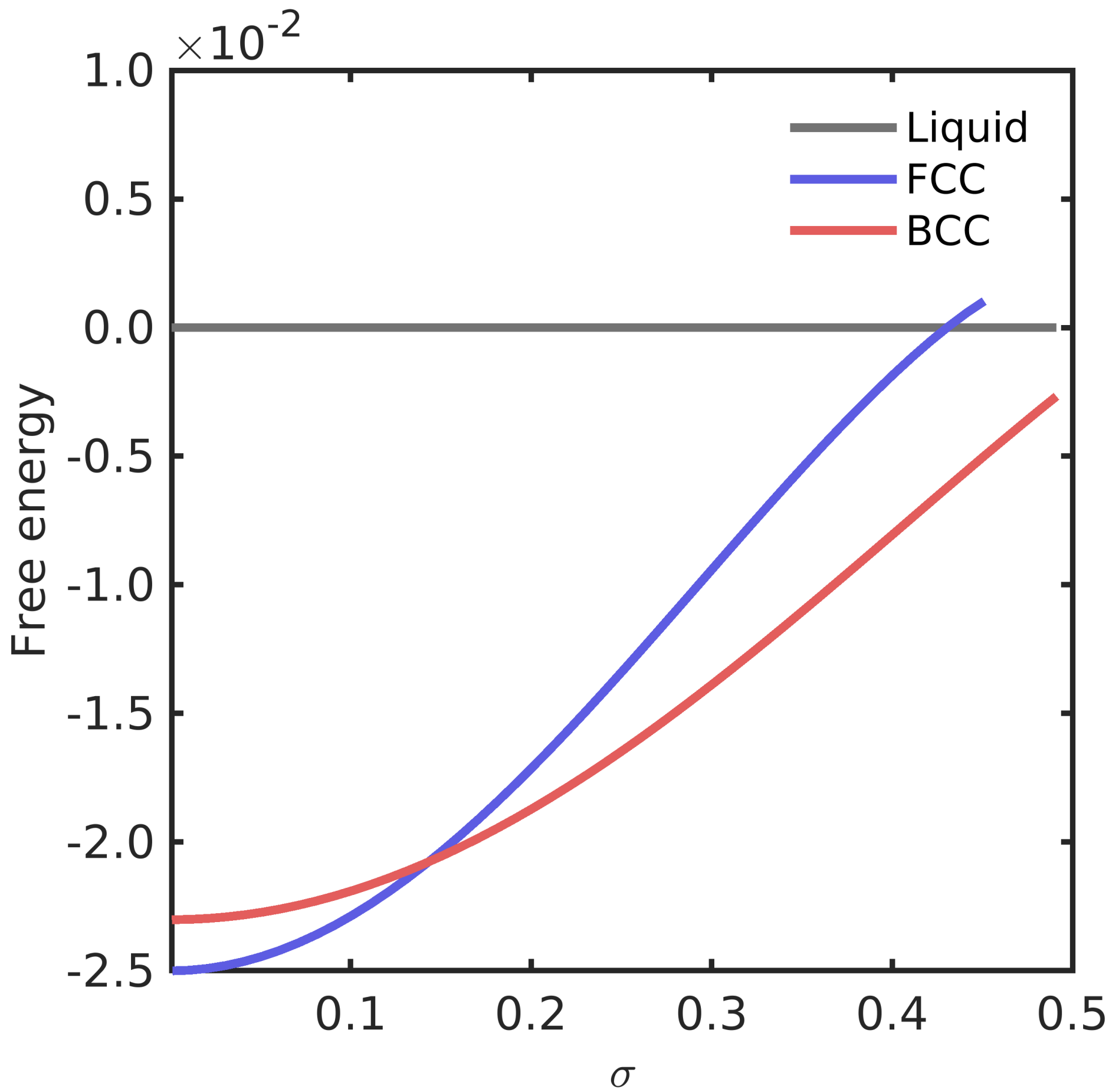

**Fig. S3. The dimensionless free energy of FCC and BCC phases with respect to the effective temperature $\sigma$.** The crossover of free energy curves is used to select the transformation condition in the PFC simulations.

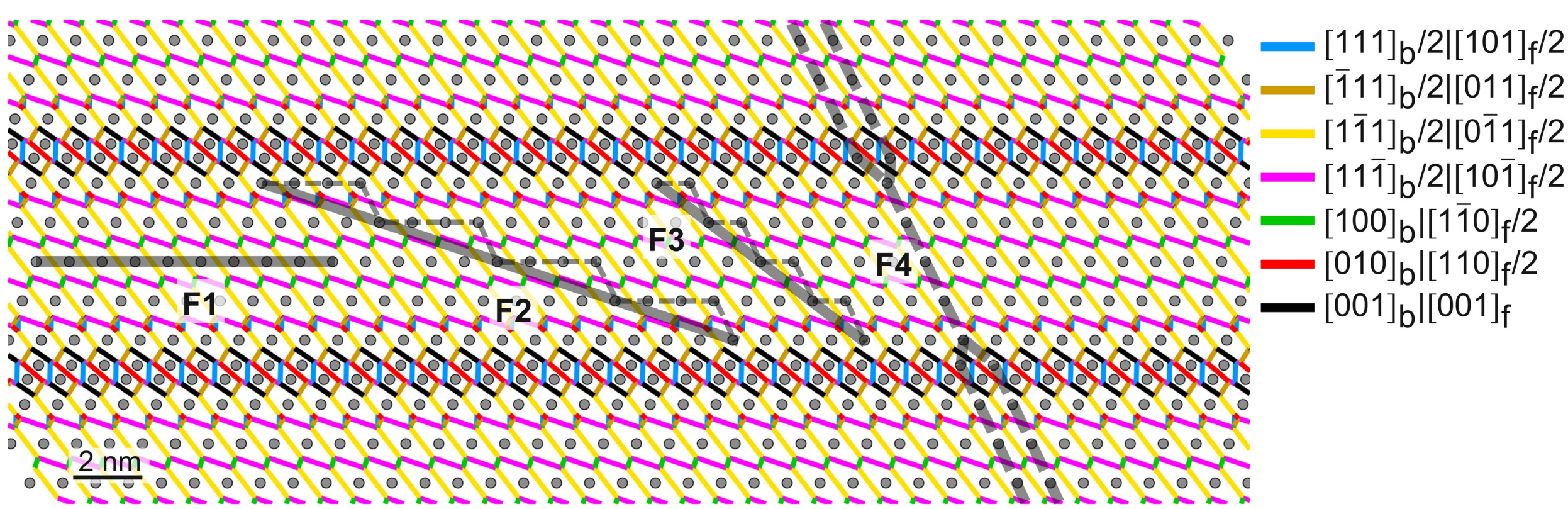

**Fig. S4. Geometric construction of the facets and interfacial defect structures of a lath-shaped Cr precipitate in a Cu matrix.** For direct comparison, the corresponding experimental TEM image and terrace–ledge schematic are shown in Figs. 1A and 2B, respectively, of Luo et al. (*25*). Gray dots are edge-on generalized O-lines. Colored lines denote the Burgers vectors of the O-cell walls (legend). Thick gray lines indicate mean facet orientations, and dashed lines on F2 and F3 indicate periodic ledges. The orientation of F4 can vary depending on the choice of coarse-spaced dislocations. The calculated coarse-spaced dislocation on F4 exhibits ledge character, as also noted experimentally (*25*).

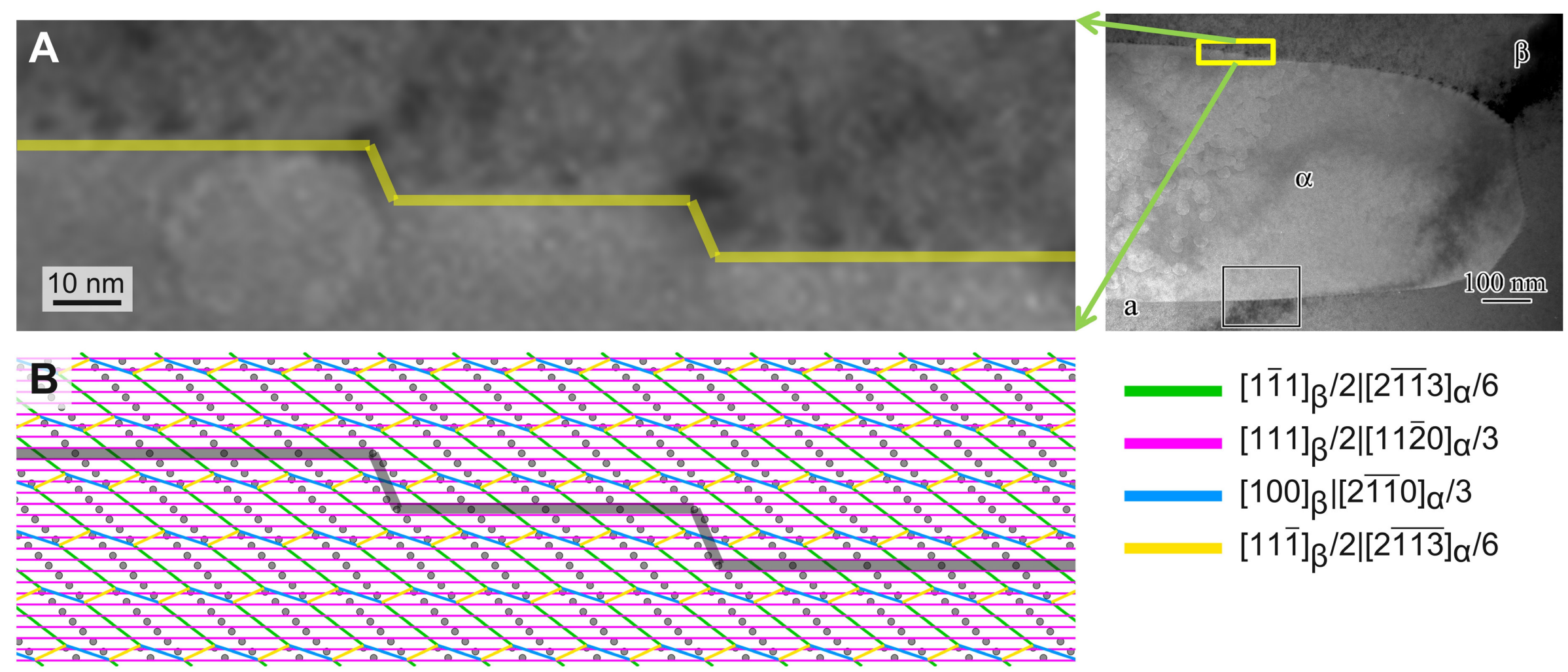

**Fig. S5. Growth-ledge structure on the HCP/BCC habit plane of an α precipitate in the β matrix (Ti–7.26 wt%Cr alloys).** (**A**) TEM image highlighting the growth ledges on the habit plane (enlarged view shown on the left; the overview image indicates the precipitate morphology and the region of interest). The ledge height is ~10 nm. (**B**) Interfacial dislocation/ledge construction based on the generalized O-element/O-cell wall calculations, illustrating the periodic dislocations on the terrace (by intersections with green O-cell walls) and the fine- and coarse-spaced dislocations associated with the ledge riser (by intersections with pink and blue O-cell walls). (A) is reproduced from the thesis of Ye (*58*).

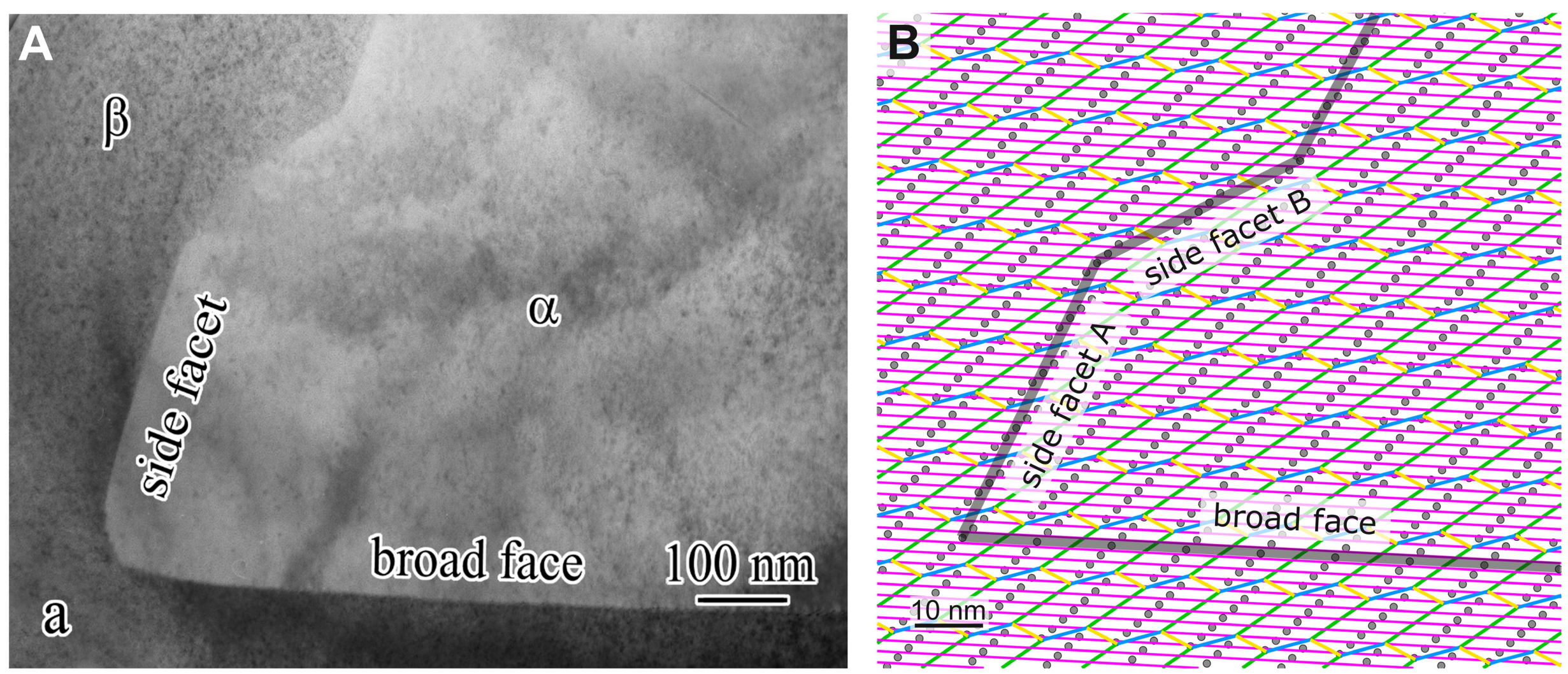

**Fig. S6. Alternative side-facet configurations of an α precipitate in Ti–7.26 wt%Cr.** (**A**) TEM image showing the broad face and side facets of an α precipitate in the β matrix. (**B**) Corresponding generalized O-element/O-cell wall construction highlighting the broad face and two possible side-facet orientations ("side facet A" and "side facet B") associated with different choices of coarse-spaced dislocations on the side facet. (A) is reproduced from Ye et al. (*59*).

**Table S1.** Crystallographic features of PFC-predicted lath precipitate and experimentally observed lath precipitates across various alloys

| | | PFC | Duplex stainless steel (*12*) | Cu–Cr alloys (*20, 25*) | Ni–Cr alloys (*24*) |
|---|---|---|---|---|---|
| OR | $a_f/a_b$ | 1.225[**] | 1.255 | 1.264 | 1.255 |
| | $\theta_{\text{p-p}}$(°) | 0 | 0.45 | 0 | 0 |
| | $\theta_{\text{d-d}}$(°) | 0 | 0.45 | 0.5 | ≤0.9 |
| Long axis | | $[0\bar{1}1]_f$ | $[0.08\ \overline{0.77}\ 0.64]_f$ | $[0.13\ \overline{0.76}\ 0.64]_f$ | $[0.06\ \overline{0.76}\ 0.65]_f$ |
| HP | $\mathbf{n}_{\text{HP}}$ | $(2\ 0.8\ 0.8)_f$ | $(2\ 1.1\ 1)_f$ | $(2\ 1.3\ 1.1)_f$ | $(2\ 1\ 1)_f$ |
| | $\mathbf{b}_{\text{HP}}$ | $\mathbf{b}_1$ | $\mathbf{b}_1$[*] | - | - |
| | $d_{\text{HP}}/a_f$ | 2.4 | 2.5[*] | - | - |
| SF | $\mathbf{n}_{\text{SF}}$ | $(\overline{1.1}\ 2\ 2)_f$ | $(\overline{0.4}\ 1.6\ 2)_f$ | $(\overline{0.7}\ 2\ 2)_f$~ $(\overline{1.5}\ 2\ 2)_f$ | $(\overline{0.5}\ 1.9\ 2)_f$ |
| | $\mathbf{b}_{\text{SF-fine}}$ | $\mathbf{b}_2$ | $\mathbf{b}_2$[*] | - | - |
| | $d_{\text{SF-fine}}/a_f$ | 3.3 | 3.3[*] | 3.4~4.1 | 4.1 |
| | $\mathbf{b}_{\text{SF-coarse}}$ | - | $[110]_f/2 \vert [010]_b$ | - | - |
| | $d_{\text{SF-coarse}}/a_f$ | - | 26.5 | 20.5~23.2 | 32.9 |

$\theta_{\text{p-p}}$: Angle between $(111)_f$ and $(011)_b$; $\theta_{\text{d-d}}$: Angle between $[0\bar{1}1]_f$ and $[1\bar{1}1]_b$

OR: orientation relationship; HP: habit plane; SF: side facet;

**n**: facet normal; **b**: Burgers vector; *d*: dislocation spacing

[*] Calculated results based on geometric model

[**] Minor deviations from experiments mainly arise from the idealized lattice-parameter ratio ($a_f/a_b=\sqrt{3/2}$) adopted in the structural PFC model. This enforces an exact match between the Burgers-vector magnitudes and the close-packed plane spacings of the FCC and BCC lattices, respectively. In real alloys, small mismatches in these quantities can introduce slight angular deviations in the precipitate long axis and facet orientations, plus additional coarse-spaced misfit dislocations on side facets.

**Table S2.** Comparison between the experimental and calculated crystallographic features of the lath Cr precipitate in the Cu matrix

| | | Exp. (*20, 25*) | Cal. | Dif. |
|---|---|---|---|---|
| Long axis | | $[1\bar{6}5]_f$ | $[1.0\ \bar{6}\ 5.0]_f$ | 0.1° |
| F1 | **n** | $(5.5\ 3.5\ 3.1)_f$ | $(5.5\ 3.4\ 3.0)_f$ | 0.6° |
| F2 | **n** | $(1\ 1\ 1)_f$ | $(0.88\ 0.98\ 1)_f$ | 3.2° |
| | *d* (nm) | 3.75 | 3.74 | 0.01 |
| F3 | **n** | $(1.6\ 4.1\ 4.6)_f$ | $(1.2\ 4.0\ 4.6)_f$ | 3.9° |
| | *d* (nm) | 1.9 | 1.94 | 0.04 |
| F4 (curved) | **n** | $(\bar{1}33)_f$~<br>$(\bar{3}44)_f$ | $(\overline{1.2}\ 2.3\ 3)_f$~<br>$(\overline{2.5}\ 2.9\ 4)_f$ | 9°<br>8° |
| | *d* (nm) | Fine: 1.25~1.5<br>Coarse: 7.5~8.5 | Fine: 1.39~1.47<br>Coarse: 6.8~7.3 | |
| *h* (nm) | | ~1.1 | 1.1 | 0.0 |
| $\theta_{p\text{-}p}$ (°) | | 0 | 0.5 | |
| $\theta_{d\text{-}d}$ (°) | | 0.5 | 0.5 | |

$\theta_{p\text{-}p}$: Angle between $(111)_f$ and $(011)_b$; $\theta_{d\text{-}d}$: Angle between $[0\bar{1}1]_f$ and $[1\bar{1}1]_b$
**n**: facet normal; *d*: dislocation/ledge spacing; *h*: smallest ledge height on habit plane